\documentclass[aps,prb,reprint,amsmath,amssymb,showpacs,showkeys,superscriptaddress]{revtex4-1}

\usepackage{graphicx}
\usepackage{dcolumn}
\usepackage{bm}
\usepackage[mathlines]{lineno}

\begin{document}

\title{Magnetic properties of Sn$_{1\textrm{-}x}$Cr$_{x}$Te diluted magnetic semiconductors}

\author{L.~Kilanski}
\email[Electronic mail: ]{lukasz.kilanski@ifpan.edu.pl}
\author{A.~Podg\'{o}rni}
\author{M.~G\'{o}rska}
\author{W.~Dobrowolski}
\affiliation{Institute of Physics, Polish Academy of Sciences, al. Lotnikow 32/46, 02-668 Warsaw, Poland}

\author{V.~E.~Slynko}
\author{E.~I.~Slynko}
\affiliation{Institute of Materials Science Problems, Ukrainian Academy of Sciences, 5 Wilde Street, 274001 Chernovtsy, Ukraine}

\date{\today}

\begin{abstract}

We present the studies of Sn$_{1\textrm{-}x}$Cr$_{x}$Te semimagnetic semiconductors with chemical composition $x$ ranging from 0.004 to 0.012. The structural characterization indicates that even at low average Cr-content $x$$\,$$\leq$$\,$0.012, the aggregation into micrometer size clusters appears in our samples. The magnetic properties are affected by the presence of clusters. In all our samples we observe the transition into the ordered state at temperatures between 130 and 140$\;$K. The analysis of both static and dynamic magnetic susceptibility data indicates that the spin-glass-like state is observed in our samples. The addition of Cr to the alloy seems to shift the spin-glass-like transition from 130$\;$K for $x$$\,$$=$$\,$0.004 to 140$\;$K for $x$$\,$$=$$\,$0.012.

\end{abstract}

\keywords{semimagnetic-semiconductors; magnetic-impurity-interactions, exchange-interactions}

\pacs{72.80.Ga, 75.30.Hx, 75.30.Et, 75.50.Pp}



\maketitle


\section{Introduction}

Diluted magnetic semiconductors offer unique possibilities to independently tune their magnetic and electrical properties.\cite{Ohno1998a} The transition metal doped IV-VI semiconductors are of the interest because the itinerant ferromagnetism with high Curie temperature, $T_{C}$, as high as 200$\;$K is achievable for Ge$_{1\textrm{-}x}$Mn$_{x}$Te alloy with x $=$ 0.5.\cite{Lechner2010a} Chromium alloyed GeTe also shows transition temperatures as high as 160$\;$K for thin layers\cite{Fukuma2007a} and 60$\;$K for bulk crystals.\cite{Kilanski2011a, Kilanski2012a, Podgorni2012a} Eggenkamp et al. investigated magnetic properties of Sn$_{1\textrm{-}x}$Mn$_{x}$Te and found the Curie-Weiss temperature ranging from 3$\;$K up to 16$\;$K for Mn-content, $x$, changing from 0.03 to 0.1, respectively.\cite{Eggenkamp1994a} They found the Mn-hole exchange constant for Sn$_{1\textrm{-}x}$Mn$_{x}$Te equal to 0.1$\;$eV. \\ \indent In the present paper we started investigating the problem of alloying SnTe with relatively small Cr-content, $x$ below 0.012. Our main goal was to see whether chromium can be successfully introduced into the SnTe lattice. We would like to explore the main exchange mechanisms present in Sn$_{1\textrm{-}x}$Cr$_{x}$Te alloy. In particular we would like to address the following questions: (i) whether in the SnTe matrix the chromium ions will substitute cation positions in the crystal lattice, (ii) will the magnetic order with high Curie temperatures appear, (iii) will the magnetic properties of  Sn$_{1\textrm{-}x}$Cr$_{x}$Te be similar to those of Sn$_{1\textrm{-}x}$Mn$_{x}$Te.

\section{Basic characterization}

\begin{figure}[!h]
  \begin{center}
    \includegraphics[width = 0.45\textwidth, bb = 20 90 845 580]
    {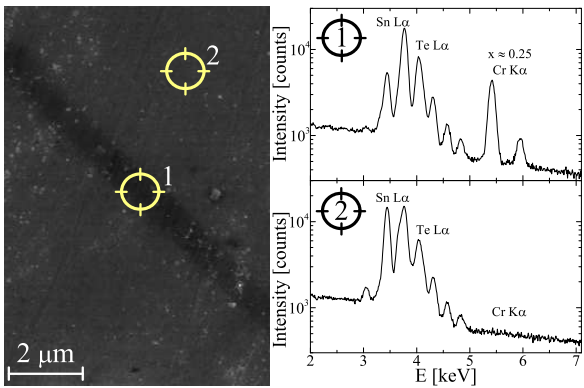}\\
  \end{center}
  \caption{\small The scanning electron microscope micrograph of the sample surface and the x-ray fluorescence spectra measured at selected spots of the selected Sn$_{0.988}$Cr$_{0.012}$Te sample.}
  \label{FigSEMEDS}
\end{figure}

The Sn$_{1\textrm{-}x}$Cr$_{x}$Te crystals being the subject of the current research were synthesized with the use of a modified Bridgman method. The modifications of the growth procedure were similar to the ones applied for the growth of alumina crystals.\cite{Aust1958a} The modifications consisted of the presence of additional heating elements creating radial temperature gradients present in the growth furnace. This improved the crystal quality and thus reduced the number of individual crystal blocks in the as grown ingot from a few down to a single one. \\ \indent The chemical composition of our samples was studied with the use of energy dispersive x-ray fluorescence technique (EDXRF). This method allows the determination of chemical composition of the alloy with maximum relative errors in the molar fraction of alloying elements, $x$, not exceeding 10\%. The results of the EDXRF measurements show that the chemical composition of our Sn$_{1\textrm{-}x}$Cr$_{x}$Te samples changes in the range of 0.004$\,$$\leq$$\,$$x$$\,$$\leq$$\,$0.012. We focus, therefore, on samples of low chromium content due to the probable low solubility of Cr in SnTe. \\ \indent The crystallographic quality of the Sn$_{1\textrm{-}x\textrm{-}y}$Cr$_{x}$Eu$_{y}$Te samples was studied with the use of a standard powder x-ray diffraction method (XRD). The XRD measurements were done at room temperature using Siemens D5000 diffractometer. The Rietveld refinement method was used in order to calculate the crystallographic parameters of our samples. The obtained XRD results indicate the presence of a single cubic NaCl phase in our samples. The lattice parameter $a$, calculated using Rietveld method is close to the value for SnTe crystals, i.e. $a$$\,$$=$$\,$6.327$\;$$\textrm{\AA}$.\cite{Nimtz1983a}  \\ \indent The Hitachi SU-70 Analytical ultra high resolution field emission scanning electron microscope (SEM) coupled with Thermo Fisher NSS 312 energy dispersive x-ray spectrometer (EDS) equipped with SDD-type detector was used in order to study chemical homogeneity of our Sn$_{1\textrm{-}x}$Cr$_{x}$Te samples. A series of SEM micrographs done for our samples indicates the presence of microscopic regions (see Fig.$\;$\ref{FigSEMEDS}) the chemical composition different than the bulk of the crystal.
Detailed measurements showed that the longitudinal precipitations of Sn$_{1\textrm{-}x}$Cr$_{x}$Te with rather high Cr-content, $x$$\,$$\approx$$\,$0.25$\pm$0.03, usually have a diameter of 1-2$\;$$\mu$m and a length of up to 10-15$\;$$\mu$m. \\ \indent A basic magnetotransport characterization of our samples was performed. We used the standard six contact dc current Hall effect technique. The Hall effect measurements were done over the temperature range from 4.3$\;$K up to 300$\;$K in the presence of a constant magnetic field of induction not exceeding $B$$\,$$=$$\,$1.5$\;$T. The measurements indicate that in our samples the resistivity $\rho_{xx}$ as a function of the temperature is typical of degenerate semiconductors. The Hall effect measurements show that our samples have $p$-type conductivity with relatively high carrier concentration $n$$\,$$\approx$$\,$2$\times$10$^{20}$$\;$cm$^{-3}$ and Hall mobility $\mu$ equal to 400$\;$cm$^{2}$/(V$\cdot$s) at $T$$\,$$=$$\,$4.3$\;$ slowly decreasing as a function of the temperature down to $\mu$$\,$$=$$\,$130$\;$cm$^{2}$/(V$\cdot$s) at $T$$\,$$=$$\,$300$\;$K. The mobility reduction with increasing the temperature is an obvious consequence of phonon scattering increase in the Sn$_{1\textrm{-}x}$Cr$_{x}$Te lattice.

\section{Magnetic properties}

\begin{figure}[h!]
  \begin{center}
    \includegraphics[width = 0.38\textwidth, bb = 0 50 590 536]
    {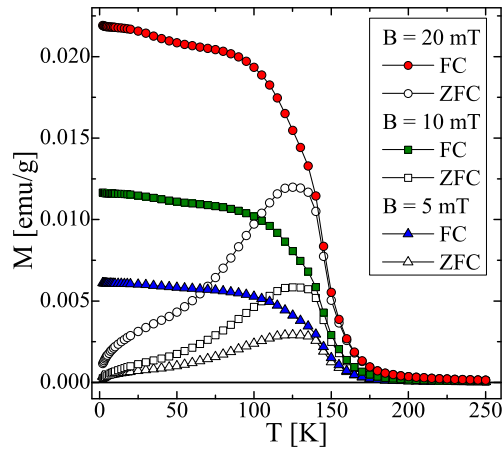}\\
  \end{center}
  \caption{\small The magnetization $M$ as a function of temperature measured at selected magnetic fields $B$ with the sample cooled in the absence (ZFC curves) and in the presence of an external magnetic field (FC curves) for the selected Sn$_{0.988}$Cr$_{0.012}$Te sample.}
  \label{FigMT}
\end{figure}

\begin{figure}[h!]
  \begin{center}
    \includegraphics[width = 0.38\textwidth, bb = 0 40 590 536]
    {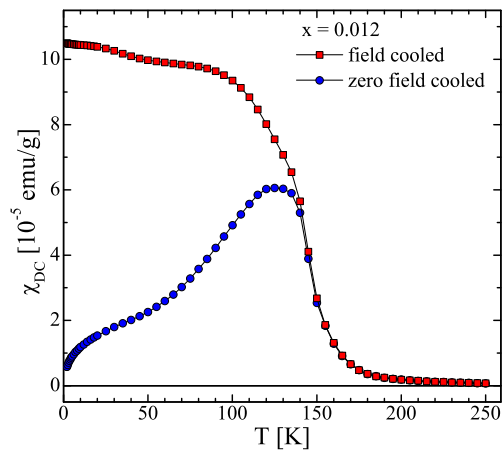}\\
  \end{center}
  \caption{\small The static magnetic susceptibility as a function of the temperature measured with sample cooled in the absence (ZFC curve) and in the presence of an external magnetic field (FC curve) for the selected Sn$_{0.988}$Cr$_{0.012}$Te sample.}
  \label{FigXdcT}
\end{figure}

\begin{figure}[h!]
  \begin{center}
    \includegraphics[width = 0.38\textwidth, bb = 0 50 590 536]
    {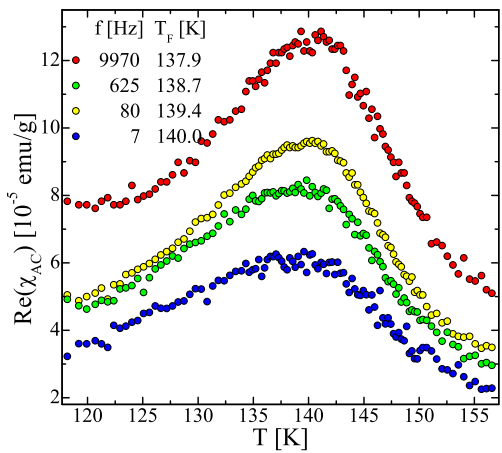}\\
  \end{center}
  \caption{\small The dynamic magnetic susceptibility as a function of the temperature measured at four selected frequencies of the alternating magnetic field for the selected Sn$_{0.988}$Cr$_{0.012}$Te sample.}
  \label{FigXacT}
\end{figure}

Magnetic properties of our samples were studied with the use of LakeShore 7229 magnetometer/susceptometer system and Quantum Design XL-5 magnetometer. \\ \indent At first, the detailed measurements of the magnetization, $M$, were performed over a wide temperature range from 2$\;$K up to 250$\;$K. During the measurements, the sample was put into the magnetic field $B$ with 3 different values 5, 10, and 20$\;$mT. The $M$($T$) measurements were performed under two conditions at which the sample cooling was performed: (i) in the presence of the external magnetic field (FC - field curves) and (ii) in the absence of an external magnetic field (ZFC - zero field curves). Corrections were made for the magnetic contribution of the sample holder. The results of the measurements are presented in Fig.$\;$\ref{FigMT}.
Our results indicate the presence of a magnetic transition in the Sn$_{1\textrm{-}x}$Cr$_{x}$Te crystals, slightly increasing as a function of Cr content, $x$, from 130$\;$ for $x$$\,$$=$$\,$0.004 up to about 140$\;$K for $x$$\,$$=$$\,$0.012. The ZFC magnetization, $M$ at, $T$$\,$$<$$\,$120$\;$K decreases with decreasing the temperature indicating, that we observe spin-glass, superparamagnetic, or possibly antiferromagnetic state in our samples. However, a more detailed interpretation of the observed data can be done using the static magnetic susceptibility. \\ \indent The dc magnetic susceptibility $\chi_{DC}$ can be  calculated as $\frac{\delta M}{\delta B}$$\big{|}_{T=const.}$ for both ZFC and FC $M$($T$) curves. The results of our calculation in a form of temperature dependencies of both ZFC and FC static susceptibility $\chi_{DC}$ are presented in Fig.$\;$\ref{FigXdcT}.
The ZFC static magnetic susceptibility $\chi_{DC}$ shows a maximum between 130 and 140$\;$K, which is a signature of a presence of a magnetic order in our samples. The shape of the $\chi_{DC}$($T$) shows the large bifurcations between the FC and ZFC curve for $T$$\,$$<$$\,$120$\;$K. It indicates that we do not observe clear ferromagnetic alignment in our samples. It is difficult to determine the type of magnetic phase only from the behavior of the magnetic susceptibility, $\chi_{DC}$, versus temperature. The problem of distinguishing between the spin glass and superparamagnetic phase transition cannot be solved by measuring of a difference between magnetic susceptibilities in ZFC and FC conditions. \\ \indent In order to determine the type of magnetic ordering in the studied material, the measurements of the ac magnetic susceptibility as a function of temperature for different magnetic field amplitudes and frequencies were performed. The dynamic magnetic properties of our Sn$_{1\textrm{-}x}$Cr$_{x}$Te samples were studied with the use of LakeShore 7229 magnetometer system. The measurements of the temperature dependence of the dynamic magnetic susceptibility $\chi_{AC}$ were done in the presence of an alternating magnetic field with four different frequencies $f$ equal to 7, 80, 625, and 9970$\;$Hz and the amplitude $H_{AC}$ equal to 20$\;$Oe for $f$$\,$$\leq$$\,$625$\;$Hz and $H_{AC}$$\,$$=$$\,$1$\;$Oe for $f$$\,$$=$$\,$9970$\;$Hz. For example, the real part of the ac magnetic susceptibility, $\chi_{AC}$, is shown in Fig.$\;$\ref{FigXacT}.
The data presented in Fig.$\;$\ref{FigXacT} indicate the slight shift of the maxima in the $\chi_{AC}$($T$) dependencies towards higher temperatures with an increase in the ac magnetic field frequency. Such behavior is a signature of either superparamagnetic or spin-glass-like state in the material. The simplest way to decide between the two above magnetic states can be done with the use of the phenomenological factor, $R$, defined by Mydosh.\cite{Mydosh1994a} The frequency shifting of the peak in the dynamic susceptibility at the temperature scale can be expressed as
\begin{equation}\label{EqMydosh}
    R = \frac{\Delta T_{F}}{T_{F} \log(\Delta f)},
\end{equation}
where $\Delta$$T_{F}$$\,$$=$$\,$$T_{F}$($f_{1}$)-$T_{F}$($f_{2}$) is the difference between freezing temperatures determined at frequencies $f_{1}$ and $f_{2}$, respectively, and $\Delta$$f$$\,$$=$$\,$$f_{2}$$-$$f_{1}$. The calculated values of $R$ for our samples are similar and close to 0.02. Such a value (smaller than $R$$\,$$=$$\,$0.1) indicates that we observe the spin-glass-like state in our samples. \\ \indent Carrier mediated ferromagnetism produces ferromagnetic alignment in Sn$_{1\textrm{-}x}$Mn$_{x}$Te crystals with $T_{C}$ about an order of magnitude lower than the spin-glass temperatures of our samples. Thus, we believe that the carrier mediated long range magnetic interactions do not play a significant role in the magnetic properties of our system.

\section{Summary}

The presence of Cr-ion rich regions (with about 20-25 mol. \% of Cr) is observed in Sn$_{1-\textrm{x}}$Cr$_{x}$Te crystals with low chromium content, $x$, ranging from 0.004 up to 0.012. The inhomogeneities are not related to any Cr$_{1-\delta}$Te phases. The magnetic properties of the alloy are dominated by the presence of inhomogeneities. The spin-glass-like state at $T$$\,$$<$$\,$130$\;$K is identified in all our samples, with transition temperature changing by about 10$\;$K with increasing the average Cr-content, $x$.

\begin{acknowledgments}
\noindent The research was supported by the Foundation for Polish Science - Homing-Plus Programme co-financed by the European Union within European Regional Development Fund.
\end{acknowledgments}

\end{document}